\documentclass[12pt,preprint]{aastex}

\newcommand{\pairhh}{HH\,135/HH\,136\ }
\newcommand{\ppairhh}{HH\,135/HH\,136}
\newcommand{\hhc}{HH\,135\ }
\newcommand{\hhhc}{HH\,135}
\newcommand{\hhs}{HH\,136\ }
\newcommand{\hhhs}{HH\,136}
\newcommand{\dc}{DCld\,290.4+01.9\ }
\newcommand{\ddc}{DCld\,290.4+01.9}
\newcommand{\iras}{IRAS\,11101$-$5829\ }
\newcommand{\iiras}{IRAS\,11101$-$5829}
\newcommand{\kms}{km~s$^{-1}$\ }
\newcommand{\kkms}{km~s$^{-1}$}
\newcommand{\msol}{M$_\odot$\ }
\newcommand{\mmsol}{M$_\odot$}
\newcommand{\lsol}{L$_\odot$\ }
\newcommand{\llsol}{L$_\odot$}

\shorttitle{Optical polarimetry of HH~135/HH~136}
\shortauthors{Rodrigues et al.}

\begin{document}


\title{Optical polarimetry of HH~135/HH~136\footnote{Based 
on observations made at the Observat\'orio do Pico dos Dias,
Brazil, operated by the Laborat\'orio Nacional de Astrof\'\i sica.}}


\author{C. V. Rodrigues\altaffilmark{2}, G. R. Hickel\altaffilmark{3},
A. H. Cerqueira\altaffilmark{4}, C. G. Targon\altaffilmark{2}}


\altaffiltext{2}{Instituto Nacional de Pesquisas Espaciais/MCT --
Av. dos Astronautas, 1758 -- 12227-010 - S\~ao Jos\'e dos
Campos - SP -- Brazil -- E-mail: claudiavr@das.inpe.br}

\altaffiltext{3}{IP\&D - Universidade do Vale do Para\'\i ba -- 
Av. Shishima Hifumi, 2911 --
12244-000 - S\~ao Jos\'e dos Campos - SP -- Brazil}

\altaffiltext{4}{LATO-DCET/Universidade Estadual de Santa Cruz --
Rodovia Ilh\'eus-Itabuna, km 16 -- 45662-000 - Ilh\'eus - BA --
Brazil}


\begin{abstract}

We present optical linear polarimetry in the line of sight to \ppairhh. 
The polarimetry of the field stars reveals two populations:
one corresponds to a foreground interstellar component; 
the other originates in the interstellar medium in the vicinity of the Herbig-Haro
pair and, therefore, can be used to study the magnetic field in the
star forming region.
Its direction is aligned with the jet of \ppairhh, which 
could be an indication that the interstellar magnetic field is 
important in the outflow collimation. The interstellar magnetic field 
magnitude was estimated to be of order 90 $\mu$G. According to
recent numerical simulations, an
interstellar magnetic field of such strength can be important in the definition
of the outflow direction. There is also evidence that 
the associated dark cloud has an elongation 
parallel to the magnetic field. Our image polarimetry 
of the extended emission associated with \pairhh shows a
centro-symmetric pattern pointing to the
knot E of HH~136. Previous near infrared polarimetry
traces a different illumination center, namely \iras - the
probable exciting source of the system. This discrepancy
can be explained if the YSO emission is completely blocked
in optical wavelengths and the dominant optical source in the region
is the knot E, whose nature is uncertain. A discussion of the 
spectral energy distributions of \hhhs-E and
\iras is presented.

\end{abstract}





\section{Introduction}

Magnetic fields are believed to play a crucial role in the physics
of jets and outflows in young stellar objects (YSOs).  
The models presently proposed to explain the
outflow engine in low mass YSOs rely on a magneto-centrifugally driven mechanism
(\citealt{sha06} and references therein; \citealt{fer06}).  Whether 
the magnetic field also defines the launching mechanism and
properties of jets in high mass YSOs is still unclear. 
Some observational
findings suggest that the formation of intermediate- to high-mass stars
also proceeds via disk accretion as in its low-mass counterparts, powering
similarly highly collimated outflows \citep{mar93,bro03,dav04,gre06}.
On the other hand, the interstellar (IS) magnetic field
can be relevant in the maintenance of jets, as is suggested by the
simulations of \citet{dec05}. From an observational perspective,
\citet{men04},
based on a sample of classical T Tauri stars, suggested  that  the
objects with bright and/or long jets might have their disk axes parallel
to the interstellar magnetic field.

\object{HH 135} and \object{HH 136} are very luminous Herbig-Haro (HH) objects 
discovered by \citet{ogu92}, who presented optical imaging and
spectroscopy of the sources. They are located in Eastern Carina
in the southwest portion of the dark cloud
\object{DCld~290.4+01.9} \citep{har86} near the bright-rimmed HII region
BBW 47 \citep{bra86}. The recently discovered infrared
cluster  \#59 from \citet{dut03} is also coincident with the HH pair.
A more complete description
of the optical objects in this region is presented by \citet{ogu92}. 
The estimated distance to the optical/infrared objects in this region is in the
2.7 -- 2.9 kpc  range.
\ddc, which has a size of 28\arcmin $\times$ 12\arcmin, is included in the
CO(J=1-0) catalogue of \citet{otr00}. This line has a well defined Gaussian
shape with a FWHM of 1.0~\kms and $V_{LSR}$ = -19.8\kkms.

Infrared observation were recently used to detect
and study the physical properties of the ${\rm H_2}$ jet \citep{gre06}.
His ${\rm H_2}$ and [Fe {\sc ii}] continuum-subtracted narrow band images 
nicely trace the line emission morphology in the outflow.
A CO molecular outflow is also present
\citep{ogu98}.  \citet{chr06} present imaging circular polarimetry of
\pairhh which suggests a helical magnetic field in the outflow.

The emission knots of \hhc and \hhs are distributed in a practically
straight line \citep{ogu92}. This could be interpreted as an evidence for two jets
with a common origin. However, both jets are dominated by
blue-shifted components, which has led \citet{ogu92} to propose that
each HH object had a different source.
Subsequent infrared polarimetry of the extended 
emission associated with the HH objects has
shown that they have a common illuminating source, namely \object{IRAS
11101$-$5829} \citep{tam97}. The apparent contradiction of these
two observations can be avoided by the scenario proposed by 
\citet{ogu98}. In this picture, the \hhs jet is deflected by a molecular
cloud, changing from a red-shifted jet near the IRAS source to a 
blue-shifted one in its extremity (see Figure 5 of \citealt{ogu98}).

\iras is a luminous ($10^4 L_\odot$)
YSO  \citep{ogu92} associated with molecular masers of different species
\citep{bra89,tel96,wal97}. \citet{tam97} suggested that it is a Herbig
Ae/Be stars encircled by a dust disk.  In particular, the presence of
a 6.7GHz methanol maser points to a high-mass YSO \citep{wal97}. The
masers profiles have $V_{LSR}$ in the range -24 to -21 \kkms, indicating
a kinematic distance of  approximately 3 kpc. This velocity is similar to that
of the \ddc, which suggests that the IRAS source and the dark cloud are associated.
Molecular emission in CO
and CS are reported by \citet{zin95}, \citet{bro96}, and \citet{ogu98}.
The millimetric continuum image of this source  
shows evidence of more than one emission
core \citep{hil05}. These data also indicate a total cloud mass of 
230~\mmsol, consistent with the mass estimated by
\citet{ogu98} of 150~\msol using CO observations.

In this work, we present a study of the magnetic field in the interstellar
medium (ISM) around the pair \pairhh using polarimetric optical data.
Polarimetry of the optical nebula associated with the \pairhh is
also obtained. A brief discussion of the \iras and \hhhs-6 sources is presented.
In Section \ref{sec_obs}, we describe the polarimetric data and
their reduction. The results and discussion are presented in Section
\ref{sec-res}.  In the last section, we summarize our findings.

\section{Observations and data reduction}
\label{sec_obs}


The observations have been done in 2005 February 12 with the 0.60-m
Boller \& Chivens telescope at the Observat\'orio do Pico dos Dias,
Brazil, operated by the Laborat\'orio Nacional de Astrof\'\i sica,
Brazil, using a CCD camera modified by the polarimetric module described
in \citet{mag96}.  The employed technique eliminates the sky polarization
\citep{pii73,mag96}.  The CCD array used was a SITe back-illuminated,
$1024 \times 1024$ pixels. The above telescope and instrumentation gives
a field-of-view of 10\farcm5 $\times$ 10\farcm5 (1 pixel = 0\farcs62). 
The data have been taken with
an $R_C$ filter. Polarimetric standards stars
\citep{ser75,bas88,tur90} have been observed in order to calibrate the system and estimate
the instrumental polarization. The measured values of the unpolarized
standard stars were consistent with zero within the errors.
Measurements using a Glan filter
were also performed to estimate the efficiency of the instrument. They
indicate that no instrumental correction is needed.

The reduction has been done using using the
IRAF\footnote{IRAF is distributed by National Optical Astronomy
Observatories, which is operated by the Association of Universities for
Research in Astronomy, Inc., under contract with the National Science
Foundation.} facility. The images have been corrected for bias and flat-field.
Counts in the ordinary and extraordinary images
of each object were used to calculate the polarization using the method
described in \citet{mag84}.  We have utilized the {\sc IRAF} package
{\sc PCCDPACK} \citep{per00} in the polarimetric analysis.
We obtained the polarimetry of around 1\,600 objects in the field-of-view.
The results are presented and discussed in the following section.

The ordinary and extraordinary images of the extended emission
associated with \pairhh 
did not overlap allowing image polarimetry to be performed.
It has been done considering circular apertures of 2 pixels 
($\approx$1\farcs2) radius centered in points distant from 
each other by 4 pixels ($\approx$2\farcs5) in each CCD
direction. The results are presented in Section \ref{sec-res}.

In addition, we have performed differential photometry using as calibrators
USNO objects in the image: they are 490 in total.  With this we could
estimate R magnitudes for all objects in the field.

\section{Results and discussion}
\label{sec-res}

\subsection{Magnetic field geometry}
\label{sec-b-geom}

The direction of the magnetic field component in the plane of the sky
can be traced by the position angle of the optical polarization. It is valid
if one assumes that the polarization originates from the dicroic absorption of
the star light by non-spherical interstellar grains aligned by the
superparamagnetic mechanism  (\citealt{dav51}; \citealt{pur71};
a recent review on grain alignment can been found in \citealt{laz03}).

\begin{figure}
\plotone{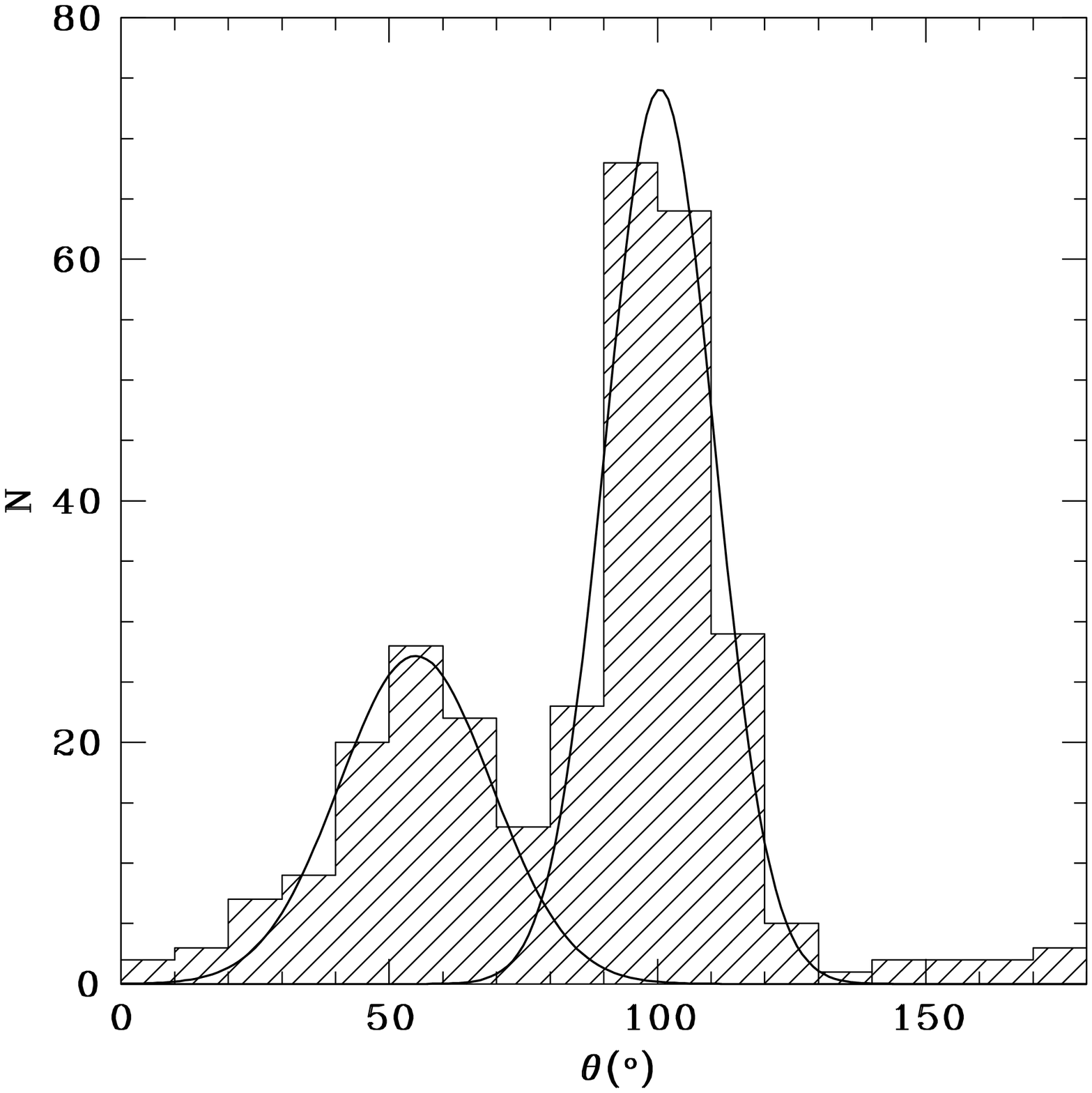}
\caption{Histogram of the position angle of the observed polarization for
fields stars with $P/\sigma_P > 5$ in the line-of-sight to \ppairhh.
A two-Gaussian fit is shown,
whose parameters can be found in Table \ref{tab-gauss}.}
\label{fig_histo_all}
\end{figure}

Figure \ref{fig_histo_all} shows the number distribution of the position
angle of polarization, $\theta$, for objects with $P/\sigma_P > 5$,
which corresponds to $\sigma_\theta < 5\fdg7$. 
We have also discarded objects that have positions superposed to
the outflow. Using these
restrictions, we reduce our sample to 303 objects. 
The distribution is clearly bimodal with peaks at approximately
55\degr\ and 100\degr. Therefore, we performed  a
two-Gaussian fit which is also shown in Figure \ref{fig_histo_all}. The
fitted parameters and errors for bins of 10\degr\ are shown in Table \ref{tab-gauss}
(lines 1 and 3). The results are statistically the same for smaller or larger
bin widths.  An inspection of the data shows that these two populations
have distinct spatial distributions and polarization magnitudes. This is
illustrated in Figures \ref{fig_small} and \ref{fig_largep} in which we
have respectively plotted the results for objects with polarization modulus
smaller and larger
than $1.5\%$ (an arbitrarily chosen number). The objects with small
values of polarization tend to be distributed in regions where the
extinction is less pronounced (Figure \ref{fig_small} - left). 
In the right panel of Figure \ref{fig_small}, we show the histogram
of the position angle of this sub-sample as well as a Gaussian curve with the
same mean
and dispersion of that of Figure \ref{fig_histo_all} centered at ~100\degr.
The agreement indicates that the population responsible for this peak in
Figure \ref{fig_histo_all} is well represented by polarization magnitudes
smaller
than 1.5\%. The large vectors  tend to be  located
in a strip running from the southwest to the northeast of the image which 
roughly corresponds to the dark cloud
(Figure \ref{fig_largep} - left). Again, one of the Gaussian curves in 
Figure \ref{fig_histo_all} fits well the distribution of position angles.

Our interpretation of the above results is that the population with smaller values of
polarization corresponds to foreground objects in the line of sight to the
HH pair, while the more polarized objects have their polarization
produced by grains associated with \ddc,
hence tracing the magnetic field alignment in the star forming region itself. To
test this hypothesis, we have used the compilation of \citet{hei00} to verify
the polarization behavior in a larger field-of-view.  We selected the
objects within a 5\degr $\times$ 5\degr\ field centered at HH~135 and with
$P/\sigma_P > 3$ (85 objects). The number distribution of the position angles and a
Gaussian fit are shown in Figure \ref{fig_heiles}. The Gaussian parameters
are presented in Table \ref{tab-gauss} (fourth line).  In spite of the larger
dispersion, the mean position angle of Heiles' objects compares
well with that of our suggested foreground component.  The mean polarization magnitude of
Heiles' objects is 1\%, which is also consistent with our data.
These results corroborate the supposition that the population having a
mean angle around 100\degr\ corresponds to the large scale,  and probably
foreground, ISM.

Another way to constrain the origin of each population is to determine
the behavior of the polarization with distance, which, however, cannot be
properly estimated with our data. From a statistical point of view, a faint object
is generally farther than a bright one. So, an alternative, despite limited, approach is
to check the polarization dependence with magnitude.
Figure \ref{fig_pol_mag}
shows that the polarization tends to increase with magnitude. This
corroborates our hypothesis that the larger polarization values are
associated with objects at larger distances.

The above discussion gives us confidence that the small polarization
component is associated with the foreground ISM in the direction of the
HH pair. Consequently, we should subtract this component from the
observations to obtain the IS
polarization produced by aligned dust {\it in} the star forming region. 
To estimate a value to the foreground component we have averaged the
polarization of the objects with the observed polarization smaller than
1.5\%. This totalizes 212 objects with a mean polarization of
0.59 $\pm$ 0.36\% @ 93.4\degr\ (the quoted error is the standard deviation of the
distribution). This
value was subtracted from our sample of 303 objects. (We would like to note that
all the arithmetics has been done using the Stokes parameters Q and U).
The number distribution of $\theta$ for objects with $P/\sigma_P > 3$
is plotted in Figure \ref{fig_histo_foreg}. The parameters of the Gaussian
fit is shown in the second line of Table  \ref{tab-gauss}. This distribution, which should represent the
magnetic field orientation in the \pairhh region, is similar
to the uncorrected, but not the same. The mean position angle
is at 41\fdg9 $\pm$ 1\fdg2.

The direction of the interstellar magnetic field found above can be
compared with the geometry of the young stellar object, in particular, with the outflow
direction. The jet position angle (from North to East - equatorial coordinates) 
has been estimated using the line joining IRAS~11101$-$5829 and a given line emitting knot.
They are: HH\,135, HH\,136/A-B-D-H. There are other knots, but whose emission occurs
mainly in the continuum, so they could not trace the jet. The
adopted position angle  for \hhs is the average of its four knots.
The resulting position angles are: $40\fdg0$ for \hhhc; and $37\fdg9~\pm~0\fdg2$
for \hhhs. Therefore, the component of the interstellar magnetic field in the
plane of sky ($\approx$42\degr) and the YSO outflow are approximately parallel.

An interstellar magnetic field aligned with the
jet is the best configuration for the propagation of
the outflow in the ISM,  as recently
demonstrated by \cite{dec05}. These authors conducted two-dimensional
numerical simulations of clumps (which, in their models, represent
time-dependent ejection from YSOs) propagating in a magnetized ISM.
They found that jets moving parallel to the ambient magnetic
field can propagate to much longer distances when compared with those
that propagate perpendicular to the magnetic field. They claim that
this could explain the correlation found by \cite{men04}
for classical T Tauri stars; namely, the bright and long jets tend
to be parallel to  the interstellar magnetic fields. The jet associated
with \pairhh has a projected size of approximately
0.5 pc  and high luminosity, so in this
object we could be seeing the effect of a parallel IS magnetic field
keeping the jet. 
On the other hand, \citet{chr06} have found evidence of a helical 
magnetic field in the outflow of \pairhh based on infrared circular polarization, 
which can concur
to collimate the jet. The present evidences, however, cannot state
unambiguously which magnetic configuration
is predominantly acting as the main large-scale collimating
mechanism in this high-mass YSO.

The emission lines of ${\rm H_2}$ and [Fe {\sc ii}] in \pairhh indicate 
a fast, dissociative J-type shock
\citep{gre06}. It is evidenced by the different space  
distributions of  these emissions. 
In a J-type shock, the transverse (relative to the propagation direction)
magnetic field is small. So the magnetic field direction inferred from our
large-scale measurements may be similar to that in the ISM in which
the shock propagates. However, we should again recall a possible
helicoidal field in the outflow \citep{chr06}, which would
produce a C-type shock or a J-type shock with precursors. More 
observations in order to constrain the detailed shock physical
conditions - as, for instance, the ${\rm H_2}$ v=0 transitions - may be helpful in 
disantangling the magnetic field geometry in the outflow region.

We could also ask if  the geometry of \dc has some correlation with the
magnetic field. Figure \ref{fig_large_scale} shows a 0\fdg5 $\times$
0\fdg5 DSS2 Red image centered at \ddc. The lines represent
the contour plot of the flux at 100~$\mu$m from IRAS.
\pairhh can be seen
in the lower right quadrant, northeast of the HII region BBW~47.
The denser portion of the cloud - as illustrated by the obscuration
at optical wavelengths and dust emission at infrared -
seems to be elongated in the northeast-southwest direction. If this
is true, the interstellar magnetic field, the HH outflow, and the cloud
elongation are all nearly parallel. This configuration is similar to
what occurs in Lynds~1641 \citep{vrb88}.

\subsection{Magnetic field strength}
\label{sec-cf}

The strength of the magnetic field in the plane of the sky, $B$, can be
estimate using

\begin{equation}
B = \left( {4 \pi \rho} \right)^{1/2} \frac{v}{\Delta\theta_B} \ ,
\label{eq_cf}
\end{equation}

\noindent where: $\rho$ is the mass density of the ISM; $v$, the
one-dimensional turbulent velocity; and $\Delta\theta_B$, the dispersion
of the magnetic field direction. This expression was proposed by
\citet{cha53} and relies on the equipartition of turbulent kinetic and
magnetic energies and isotropy of the motions in the medium. The overall
ideia behind this formula is still accepted \citep{hei05}, notwithstanding
different effects could lead the above equation not to be the best estimate 
of the actual field: large fluctuations of the magnetic field amplitude;
acting of the nonmagnetic forces on the gas; inhomogeneity of
the interstellar material \citep{zwe96}. Recent numerical simulations of
polarimetric maps of molecular clouds indicate that this formula overestimates
the magnetic field by a factor 2
\citep{ost01, pad01,hei01,hei05,mat06}.

The value of $\Delta\theta_B$ in the star formation region
can be estimated by the standard deviation of the fitted Gaussian
to the number distribution of the position angle of the intrinsic
polarization (see second line of Table \ref{tab-gauss}). This number
is, however, an overestimate of the dispersion of the magnetic field direction since it
includes the observational error associated with the $\theta$ measurement. Following
the procedure suggested by \citet{per05}, we obtain a $\Delta\theta_B$ value
of 13\fdg3.  The turbulent velocity was considered that of the dark
cloud and measured by \citet{otr00} as 1~\kkms. A total mass density 
of $1.4\times10^{-20}$ g cm$^{-3}$ has
been estimated from the number density of ${\rm H_2}$ presented in \citet{zin95}, which
was based on CS(J=2-1) measurements near the IRAS source. Considering 
a factor of 0.5 to equation (\ref{eq_cf}) - as discussed above - we obtain an
interstellar magnetic field strength of 90 $\mu$G. However, we
would like to note that this value should be interpreted as the
order of magnitude of the field. The reason is twofold. On one
hand, the observational values used in the magnetic field calculation have their own 
uncertanties. On the other hand, the values of the mass density, magnetic field dispersion, and
turbulent velocity can be tracing different portions of the ISM.
\citet{hei05} obtained that a single estimation of $B$ with the above procedure
can be in error by a factor of 7. In addition, we would like to note
 that the above estimate of $B$ should be associated with the large scale
ISM around \ppairhh, not with the outflow region.
This value is larger than that measured in the diffuse ISM of a few
$\mu$G, but it is in the range obtained for star forming regions \citep[see,
e.g.,][]{gon90,chr94}.

Recently, \citet{mat06} have studied the alignment of outflows with magnetic
fields in clouds cores through numerical simulations. They found that
the outflow tends to be aligned with the large-scale ($>$ 5000 AU) magnetic
field if the magnetic field {\it in the core}  is larger than 80~$\mu$G. Our 
above estimate of 90~$\mu$G may be interpreted as the strength in
the dark cloud, i.e., the initial magnetic field before the collapse 
(the B$_o$ of \citealt{mat06}).  So the enhanced magnetic field in the collapsing core
that originated the YSO has probably exceeded the limiting value of
80~$\mu$G, making plausible that in this region the geometry
of the magnetic field in the original
cloud determined the  direction of the YSO outflow.

\subsection{Imaging linear polarimetry of \pairhh}
\label{imagepol}

Figure \ref{fig_imagepol} shows the imaging linear polarimetry
of the emission nebula associated with \ppairhh. The background image 
is from our data: the object is seen twice because of the beam splitting produced by
the calcite block. Only measurements with 
$P/\sigma_P > 10$ are plotted. The vectors sizes show that
the polarization can be as high as 30\%. The position angles
define a clear centro-symmetric pattern,
typical of scattering, whose center 
has been calculated and coincides with knot HH~136-E
(following the denomination of \citealt{ogu92}). 
From North to South, this knot is the
second bright source in our image. 
The centers calculated using the data over the HH~135 region or the HH~136 region 
are the same.  

HH~136-E is the brightest point in the R band image, having 
a magnitude of 14.37 mag. 
The knot B, the second brightest, has a magnitude of 14.78 mag, which
corresponds to a flux 30\% smaller than knot E. 
In both estimates we have used an aperture radius of 3 pixels (= 1\farcs8). 

Our imaging R-band polarimetry indicates HH136-E as the illuminator
center of the scattering pattern, so it is unequivocally
associated with the region.  Previous K-band polarimetry
of the same region \citep{tam97} shows also a centro-symmetric pattern, but
whose center is coincident with IRAS~11101$-$5829. The dominant
source in the infrared region is NIRS~17 (\citealt{tam97} -
see also Figure 5 of \citealt{gre06}), which is coincident with knot J.
They suggested that
the IRAS source is obscured from our view by an optically thick disk, which is
evidenced by the "polarization disk", but illuminates the associated nebula
through the optically thinner pole.

The optical depth of a dusty medium grows from infrared to optical
wavelengths. Therefore, in the R band, the disk around the IRAS source can be
optically  thick even at its pole, thus preventing any flux to escape. This
could explain why \iras is not the source of the optical light being scattered in
the nebula. However, it remains as an open question the nature of the
knot E.

HH136-E is the brightest R-band source in the outflow region 
and is associated with the infrared source NIRS~9,
whose infrared colors are consistent with a 
pre-main-sequence object \citep{tam97}. 
It has a very strong optical and infrared continuum, being
practically absent of [S {\sc ii}], ${\rm H_2}$ and 
[Fe {\sc ii}] emission \citep{gre06,ogu92}. This makes a Herbig-Haro nature
quite improbable. In spite of the suggestion from Schmidt plates that
knot E has an important H$\alpha$ emission, no
slit spectroscopy at its exact position has been done.
The spectral energy distribution (SED) of the knots \hhhs-A, \hhhs-B, \hhhs-E,
and \hhc can be done using DENIS \citep{denis} and 
2MASS \citep{cut03} data, and our
photometry. None of these sources are detected in longer
wavelengths.
The SED of knot E has a rising slope from
I to K band. A black body fit to this curve provides a
bolometric luminosity of $\approx$4~\llsol, which would correspond
to a ZAMS star of $\approx$2~\mmsol.

To explain the non-trivial radial velocity structure of 
the emission knots, and considering 
a same exciting source for \hhc and \hhs as indicated
by previous  K-band polarimetry, 
\citet{ogu98}  have proposed 
a scenario in which one of the jets coming out from the 
exciting source is deflected by a nearby molecular cloud. 
The region of zero velocity is located around knots D, E,
F, and G (Figure \ref{fig-sources}; see also Figure 7 of \citealt{ogu92}). 
In this region there is also a slightly increase of
the $^{\rm 12}$CO antenna temperature \citep{ogu98}.
Besides, there is a MSX source between
knot F and G, which could represent the point of 
collision. So another possibility to the nature of knot E would
be the point where the jet collides with the molecular
cloud. 

A spectroscopic analysis of knot E as well as
a detailed velocity study of the whole complex can probably shed some
light in what is going on in this region and the true nature of this object.

\subsection{Spectral energy distribution of IRAS 11101-5829}

Figure \ref{fig_sed} shows the SED of 
IRAS 11101-5829 based on literature data (see figure legend for
the references). 
To estimate the bolometric luminosity of \iras we integrate a
cubic spline to its SED which provides a 
value of $1.32\times10^4$~L$_\odot$ at a distance of
2.7 kpc.  This is in agreement with previous estimates from \citet{ogu92}
of $1.39\times10^4$ L$_\odot$ and  
from \citet{wal97} of $1.59\times10^4$ L$_\odot$. Both of them are based on
IRAS data,  but consider different corrections. 
The above luminosity can be used to constrain the stellar mass.  Using the
massive stars evolutionary tracks of \citet{ber96} for Z = 0.02, we estimate 
an interval of 11~--~25~\msol for the mass of the central object. The higher 
masses are obtained if
the object is very young, with a convective envelope.

The SED of an embedded source contains more information than just the
luminosity of the central object. It results from the reprocessing of the stellar
flux in the circumstellar environment. To exploit this aspect, we have used
the recently available grid of models of \citet{rob06} to reproduce the
observed SED of \iiras. We have concentrated on models whose parameters
are:

\begin{itemize}

\item $1.0\times10^4$ \lsol $<$ L $<\ 1.4\times10^4$ \lsol - see discussion above;

\item 11  \msol $<$ M $<$ 25 \msol - see discussion above;

\item i = 81\fdg4. Following \citet{ogu98}, we consider that the jets make an angle with
the plane of sky of $\approx$5\degr\ and that the disk is perpendicular to the jets. Among the inclinations
provided by \citet{rob06}, we chose this value as a good approximation to \iiras;

\item aperture = 100.000 AU. At a distance of 2.7 kpc, this corresponds to 37\arcsec. This
is the largest aperture provided by the models. We have used it in order to approximate the
the angular resolution of IRAS of $\approx$ 90\arcsec.

\end{itemize}

We have then 307 models which have been visually inspected. In doing
this, we have selected the best 41 models for which we have calculated the $\chi^2$
values. We have found that the models
\#3020025 and \#3007152 produced the smallest $\chi^2$. 
Table \ref{tab-sed} shows the parameters
of the models  - we ask the reader to see a complete description of them in
\citet{rob06}. This simple modelling provides a object with a mass of $\approx$ 13 \mmsol, 
which puts the YSO near the ZAMS, with a age of ~$10^6$ years. 

The above result is unexpected considering the presence of jets that are typical
of a younger object. To discuss that issue, we would like to initially recall 
the optical knot J (=NIRS 17).  It is located at 1\farcs8 from the \iras center position.
However the error ellipse of this source is 19 x 5\arcsec, so it includes the
optical/infrared source (see Figure \ref{fig-sources}). 
In the IR and optical, the knot emission
is dominated by a strong continuum and does not have
[S {\sc ii}] emission lines \citep{ogu92,gre06}. So a Herbig-Haro
nature appears to be ruled out.  On the other hand,
the SED presents two maxima: in $\approx$60$\mu$m and
$\approx 2\mu$m. The above modelling has implicitly assumed that 
its near IR portion is caused by the circumstellar disk emission 
in the observer direction and therefore it should have the same
center position of  the far infrared emission. 
This might not be the case for \ppairhh. 
As proposed by \citet{tam97},
a possible geometry is one in which the near infrared YSO
emission (produced in disk) is obscured from the observer direct view, but it can flow from
the pole and illuminate the nebular material in the jet region. We propose that
the near IR SED (the knot J) is the YSO reflected light
in the pole cavity, as seen in HH~46 \citep{dop78}. In such a case, if the YSO emission
was isotropic, the knot J should trace the YSO's SED. However,
we should recall that the YSO emits anisotropically. Supposing the adopted
inclination is correct, the knot J should receive the emission from a smaller
inclination which has a larger near IR component. So the SED for an
inclination of $\approx$81\degr\, as seen from a
direct view, should have smaller fluxes at these wavelengths.
This would result in models having properties of a more embedded, consequently younger,
object. To do a proper modelling it would be necessary to know the 3D configuration
of the knot J and the YSO.

\section{Conclusions}

We have presented optical linear polarimetry of \pairhh and the nearby
ISM. Our main results are listed below.

\begin{itemize}

\item The polarization of stars mapping the magnetic field in the
star forming region is nearly parallel to the Herbig Haro outflow.  We suggested
that the elongation of \dc is also parallel to the magnetic field.

\item We estimated the interstellar magnetic field strength as of order 90 $\mu$G. 

\item The interstellar magnetic field direction and magnitude are
adequate to play a role in determining the outflow direction and
keeping the jet collimation. However, a collimation by a helicoidal 
magnetic field in the jet region is not discarded.

\item The R-band nebula associated with the HH pair has a clear 
reflection pattern whose center is the knot HH136-E. It
seems to be a star, but its nature could 
not be securely determined and deserves more observational effort.

\item A simple modelling of the \iras SED indicates a mass of
13 \msol and an age
of 1 Myr, which is inconsistent with the presence of
jets in the object. A less evolved stage might be found
if the knot J is assumed to be produced by the reflection of YSO light in 
the surrounding material.

\end{itemize}

\acknowledgments

\begin{figure}
\plottwo{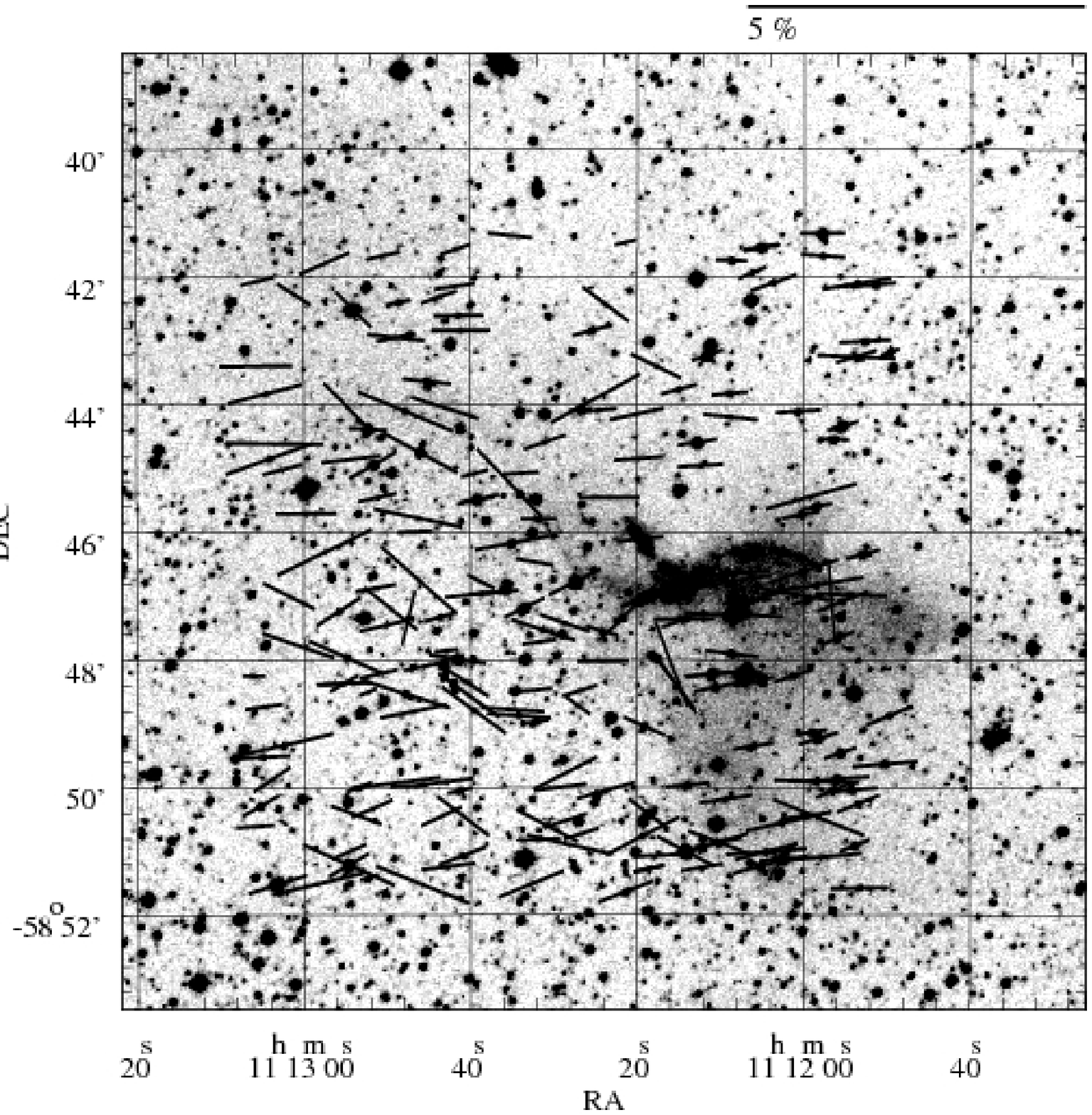}{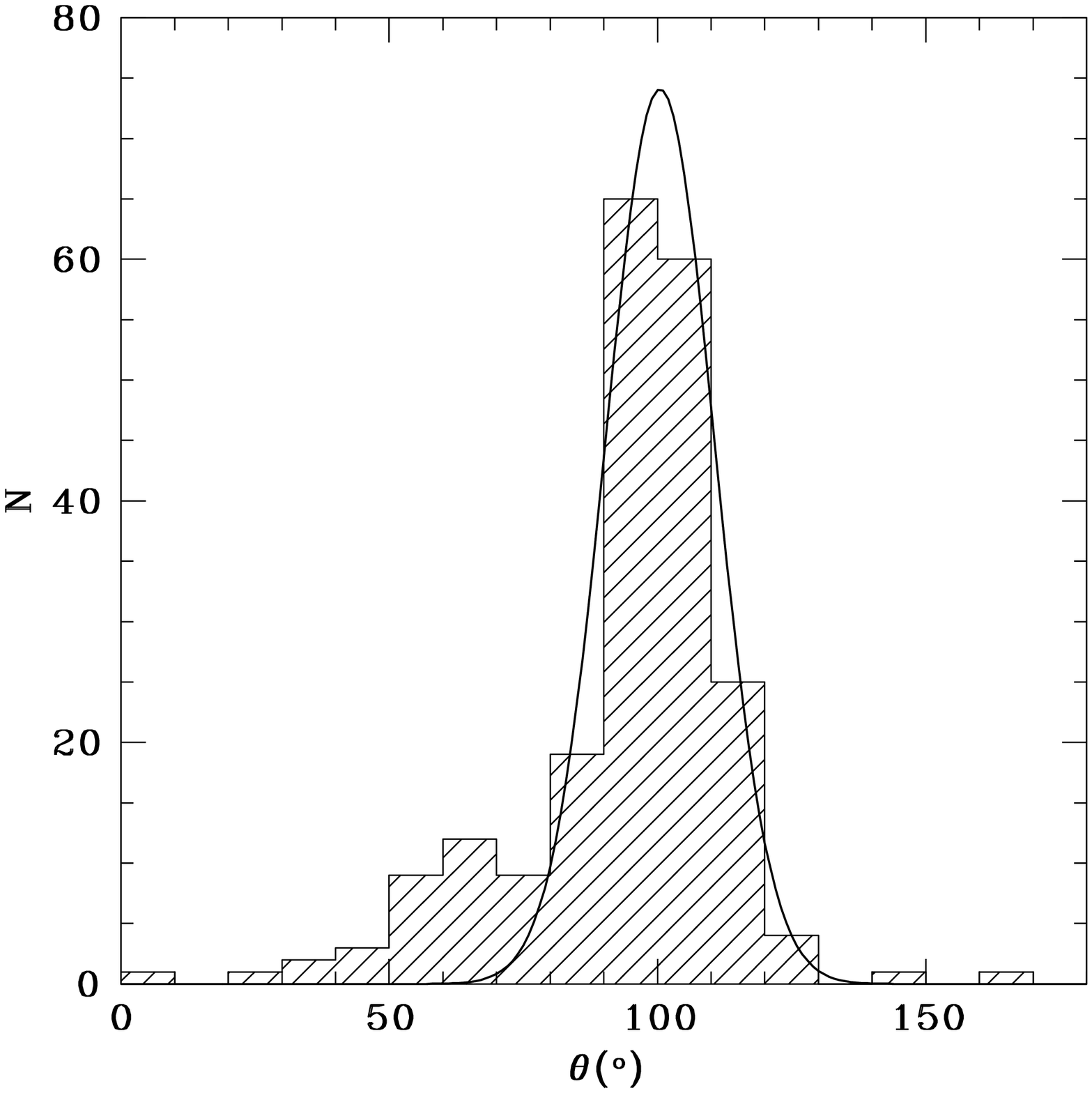}
\caption{Polarimetry of field stars in the line of sight to
HH~135/HH~136 with $P/\sigma_P > 5$ and $P < 1.5\%$.
(Left) The vectors represent
the direction and magnitude of the polarization whose scale is
presented in the upper right of the figure. 
The background image is from the DSS2/Red. 
The epoch of the coordinates is 2000.0.
(Right) Number distribution of the position angle of polarization for
the same sample. The full line is one of the Gaussian curves obtained
in the 2-Gaussian fit of Figure \ref{fig_histo_all}.}
\label{fig_small}
\end{figure}

\begin{figure}
\plottwo{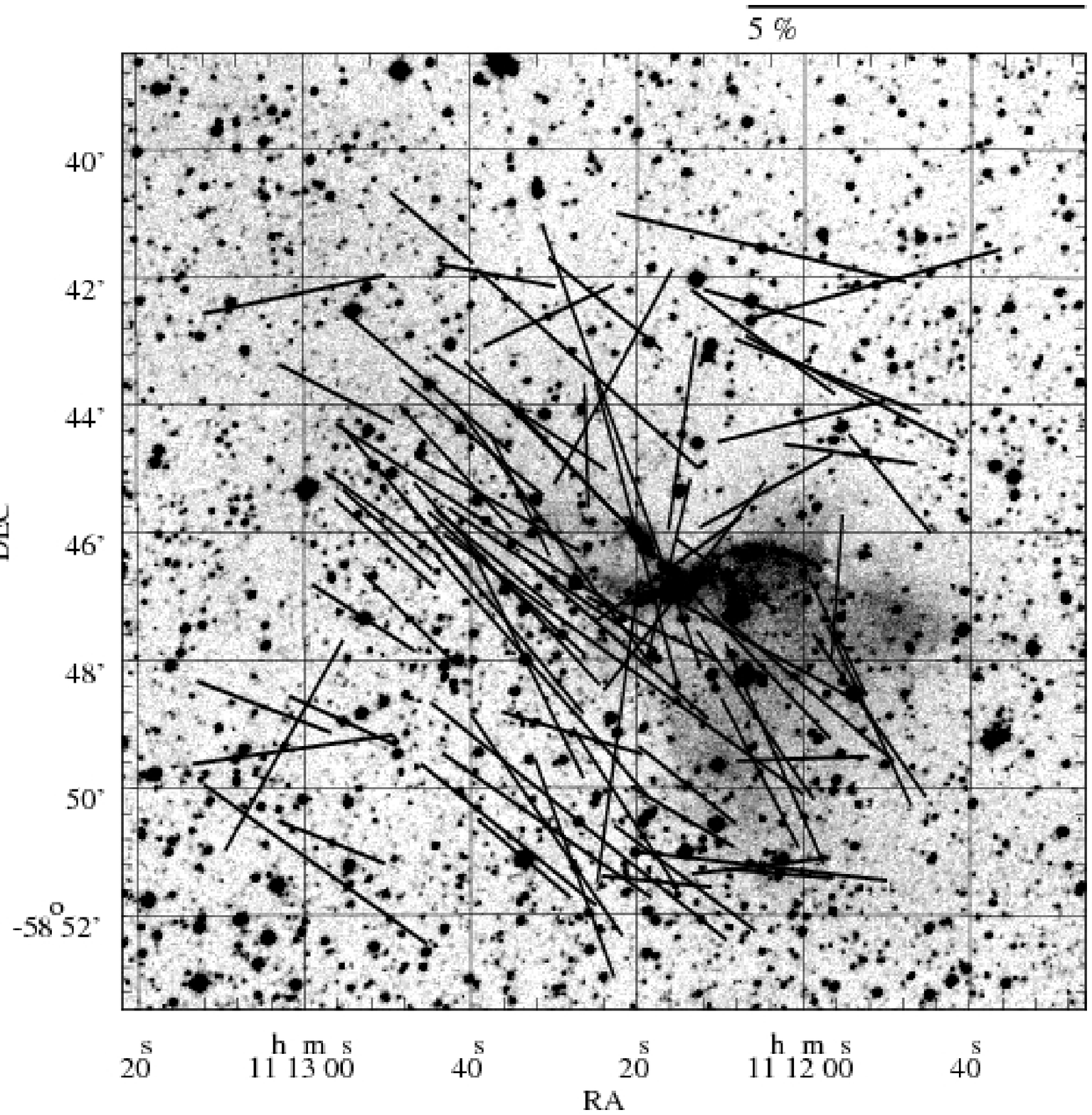}{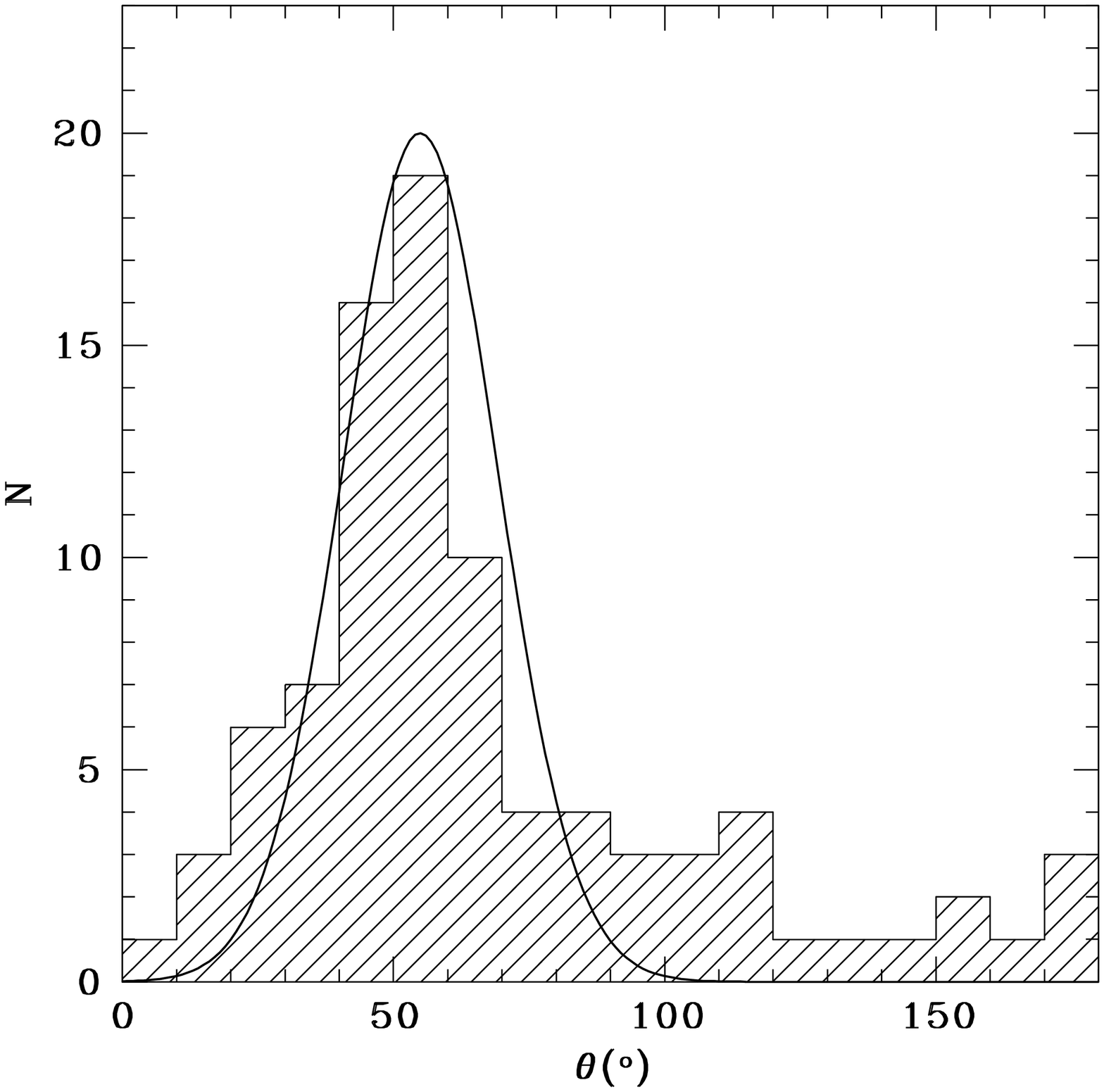}
\caption{The same as Figure \ref{fig_small} for stars with
$P/\sigma_P > 5$ and $P >1.5\%$.}
\label{fig_largep}
\end{figure}

\begin{figure}
\plotone{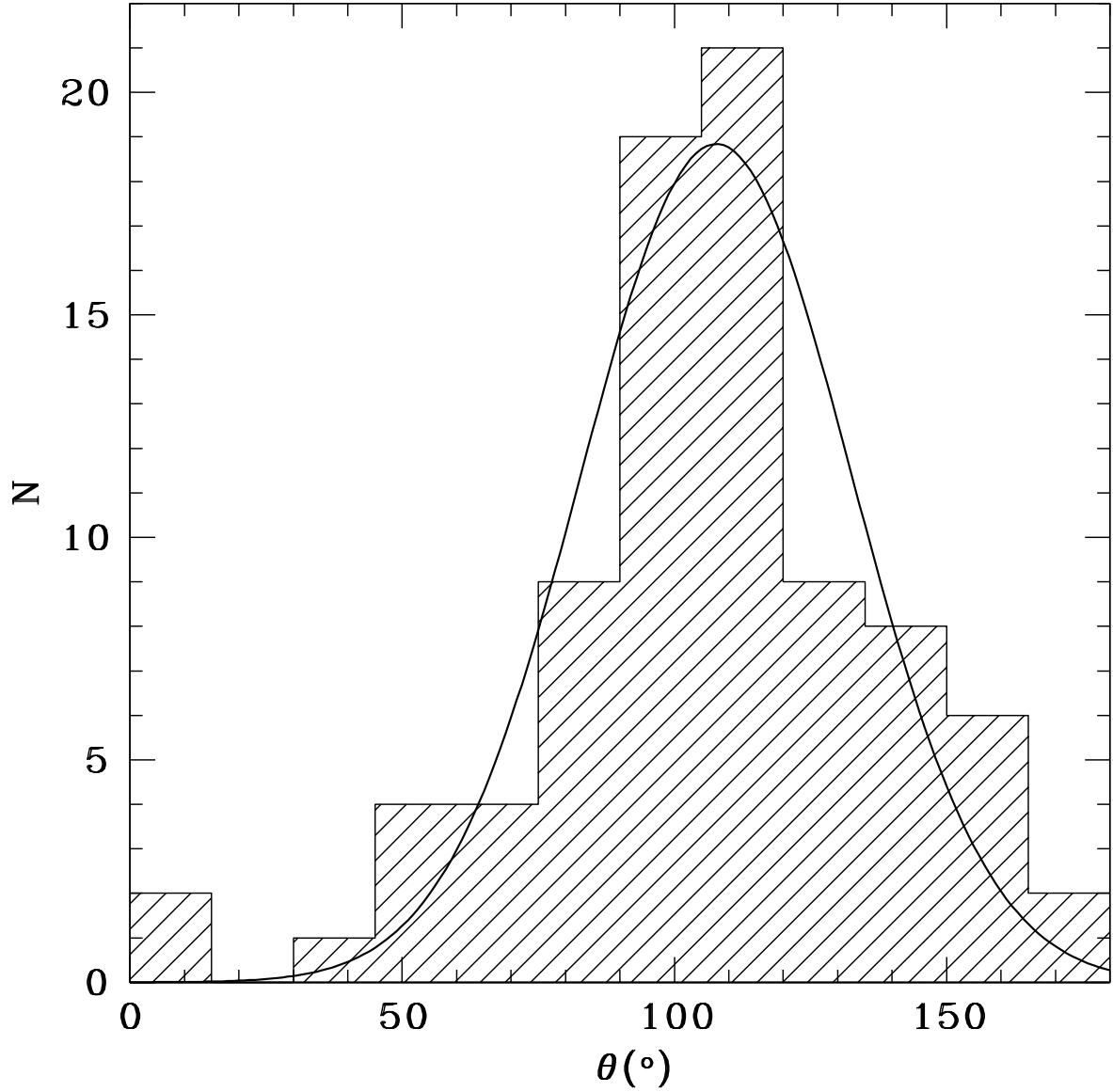}
\caption{Number distribution of position angle of polarization 
for objects in the \citet{hei00}'s catalog within a 
5\degr $\times$ 5\degr\ field-of-view centered at HH~135
and with $P/\sigma_P > 3$. A Gaussian fit is also shown - see
Table \ref{tab-gauss} for the parameters.}
\label{fig_heiles}
\end{figure}

\begin{figure}
\plotone{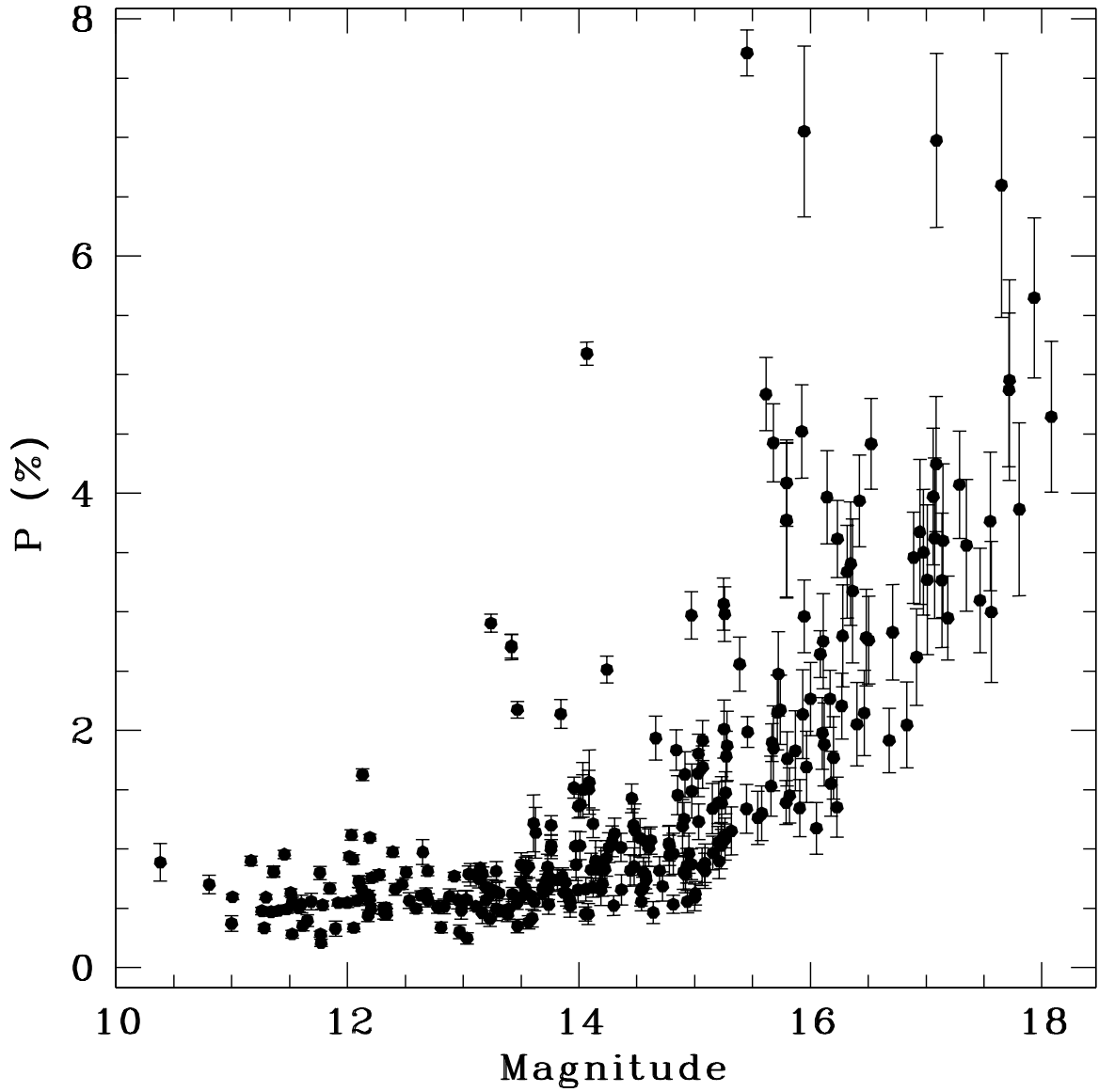}
\caption{R-band polarization versus magnitude for objects with $P/\sigma_P > 5$.}
\label{fig_pol_mag}
\end{figure}

\begin{figure}
\plotone{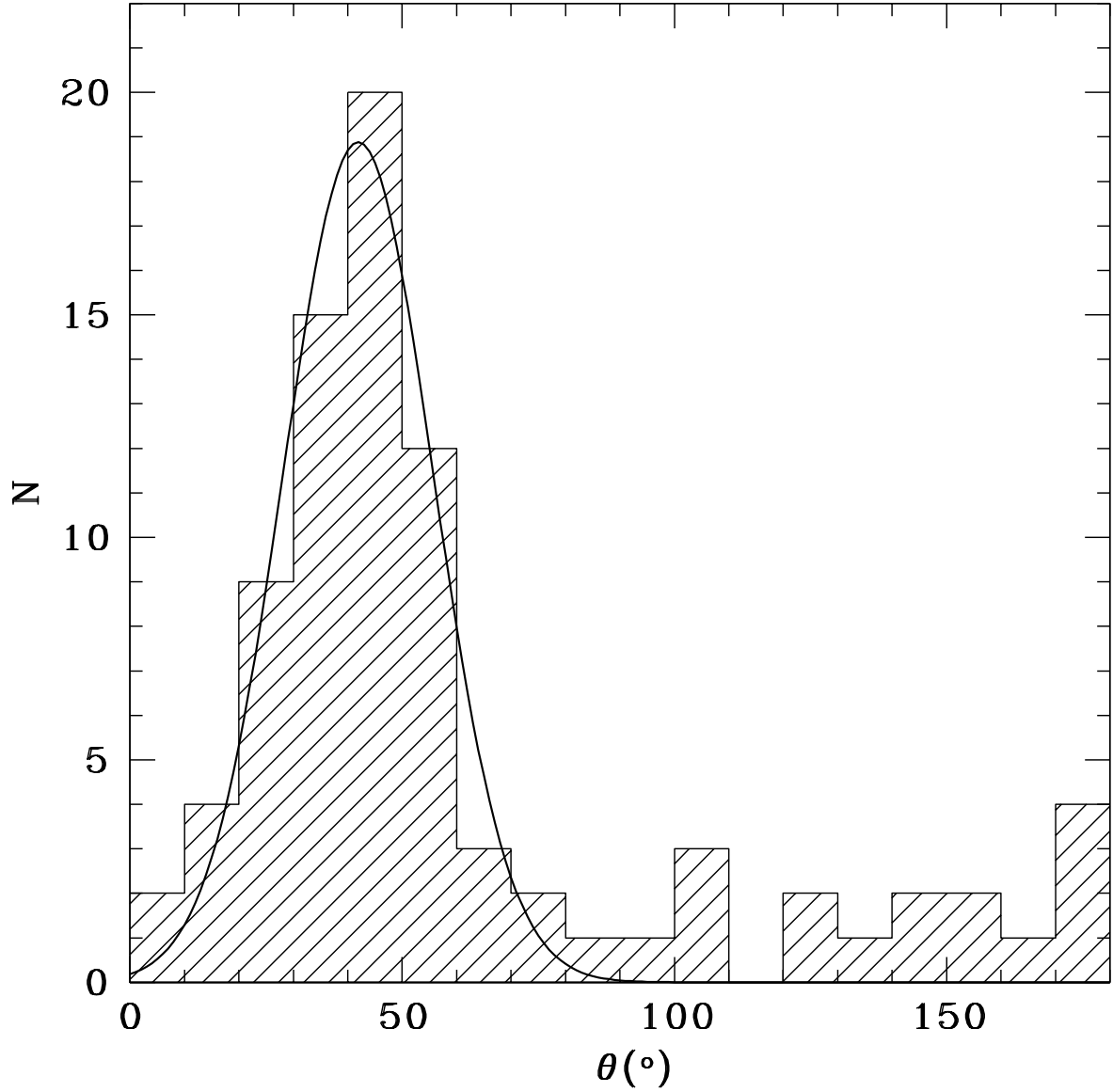}
\caption{Number distribution of position angle of intrinsic polarization 
for objects in the line-of-sight to HH~135/HH~136 with $P/\sigma_P > 3$
after the foreground polarization subtraction.
 A Gaussian fit is also shown - see
Table \ref{tab-gauss} for the parameters.}
\label{fig_histo_foreg}
\end{figure}

\begin{figure}
\plotone{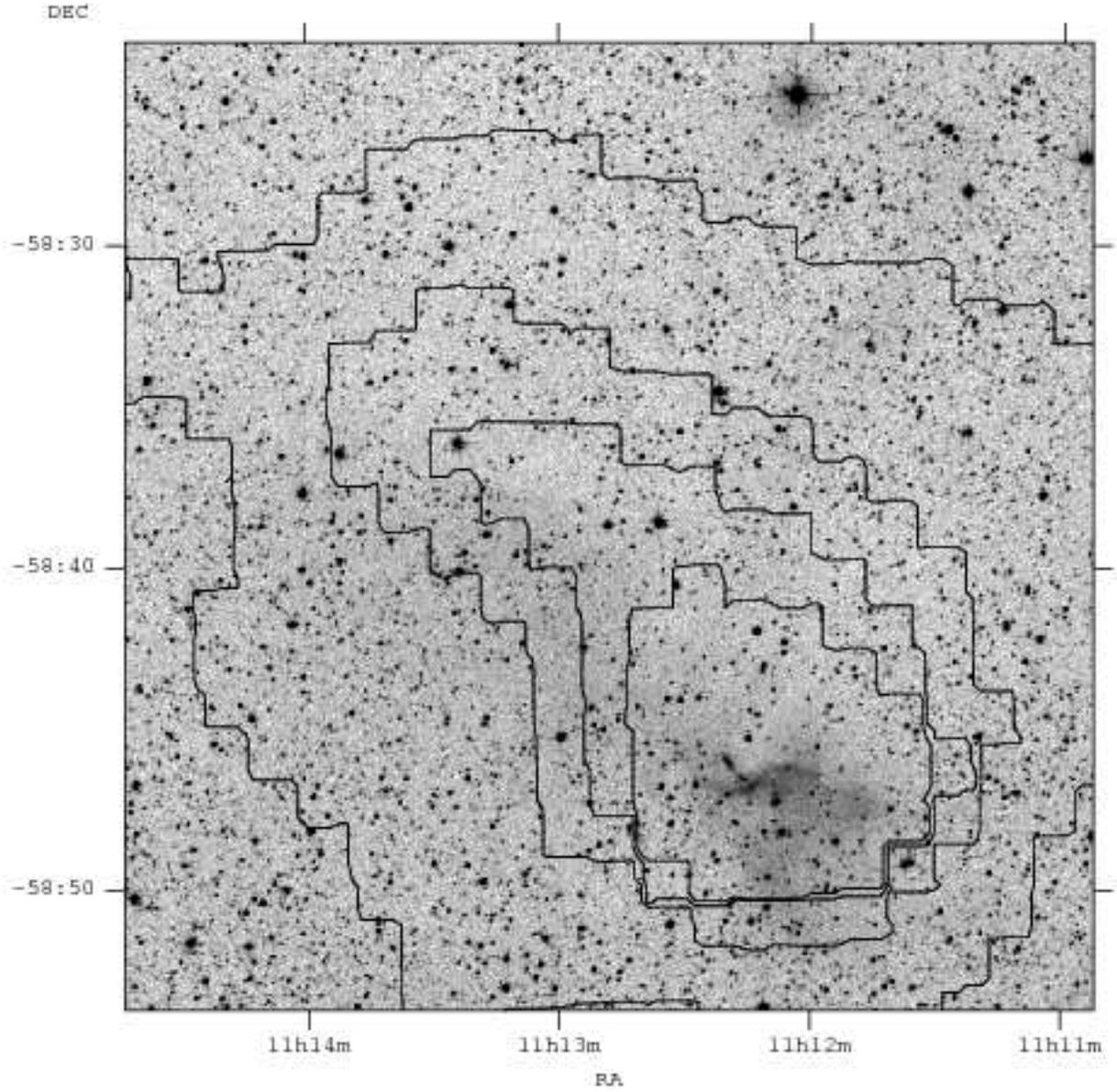}
\caption{DSS2 Red view of a 0\fdg5 $\times$ 0\fdg5 region centered at
\ddc. The contour plot of 100~$\mu$m IRAS is seen superposed.
The epoch of the coordinates is 2000.0. }
\label{fig_large_scale}
\end{figure}

\begin{figure}
\plotone{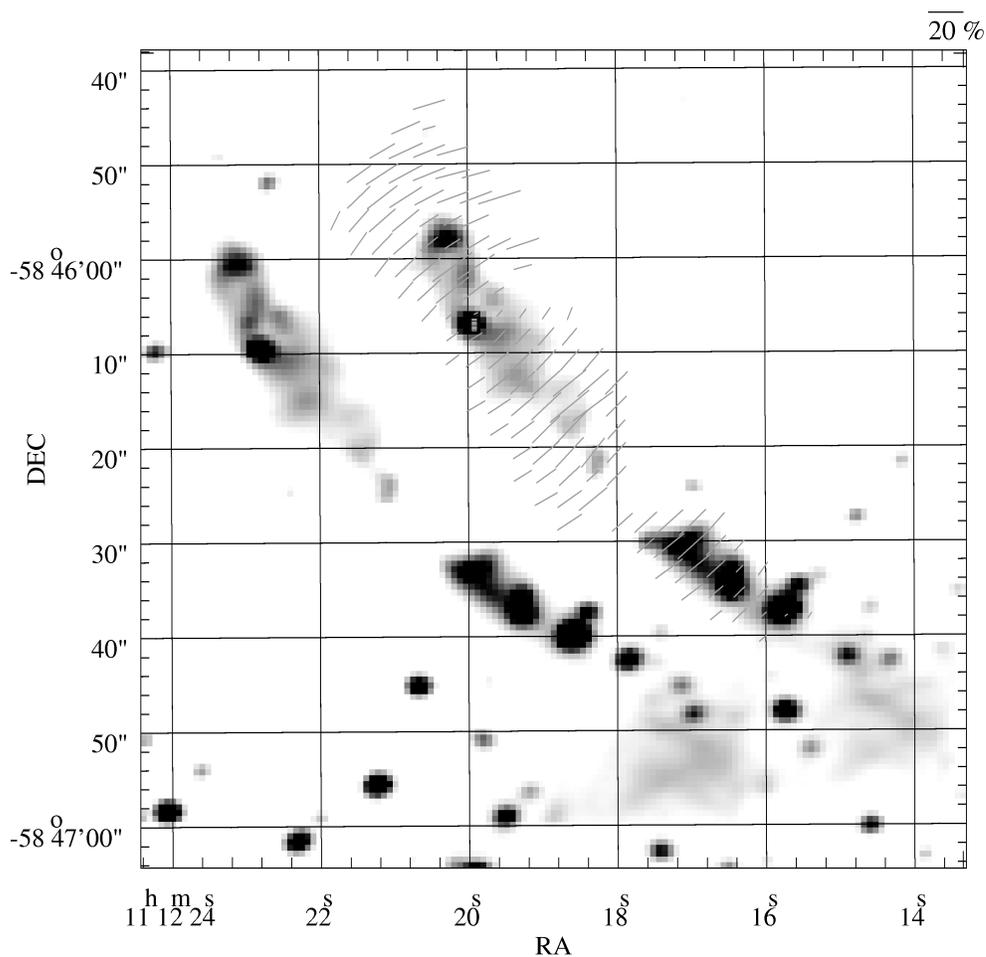}
\caption{R-band imaging polarimetry of \ppairhh. Only measurements with
$P/\sigma_P > 10$ are shown. The polarization vectors,
whose scale is presented in the top right corner, 
are superposed on our image. 
The two images correspond to the ordinary and extraordinary beams
separated by the calcite block.
The  gray retangle marks the most likely position of the ilumination source.
The coordinate scale is with respect to the right image and the vectors. The epoch is 2000.0.}
\label{fig_imagepol}
\end{figure}

\begin{figure}
\plotone{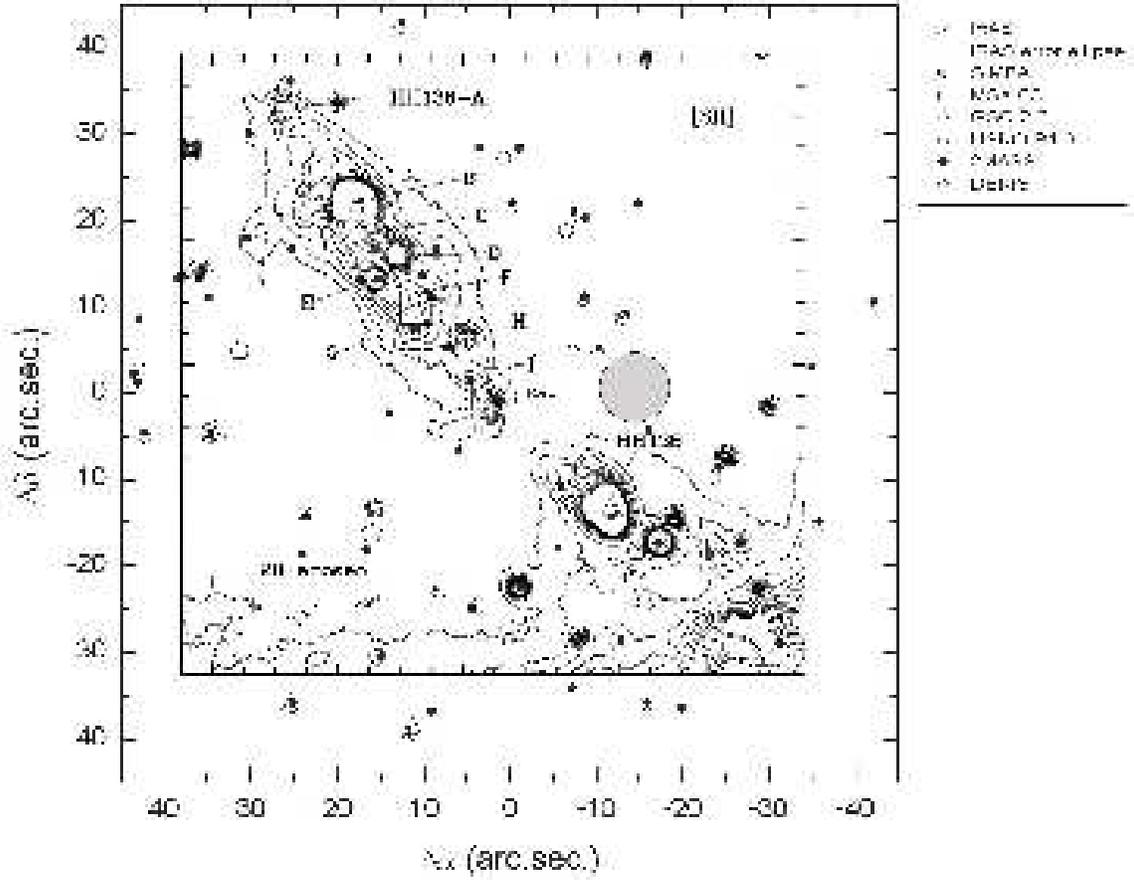}
\caption{Different wavelengths sources in the region of \ppairhh.
The contour plot is the [SII] image (continuum included) from 
\citet{gre06}. We use different symbols to represent the data origin:
see legend at left.}
\label{fig-sources}
\end{figure}

\begin{figure}
\plotone{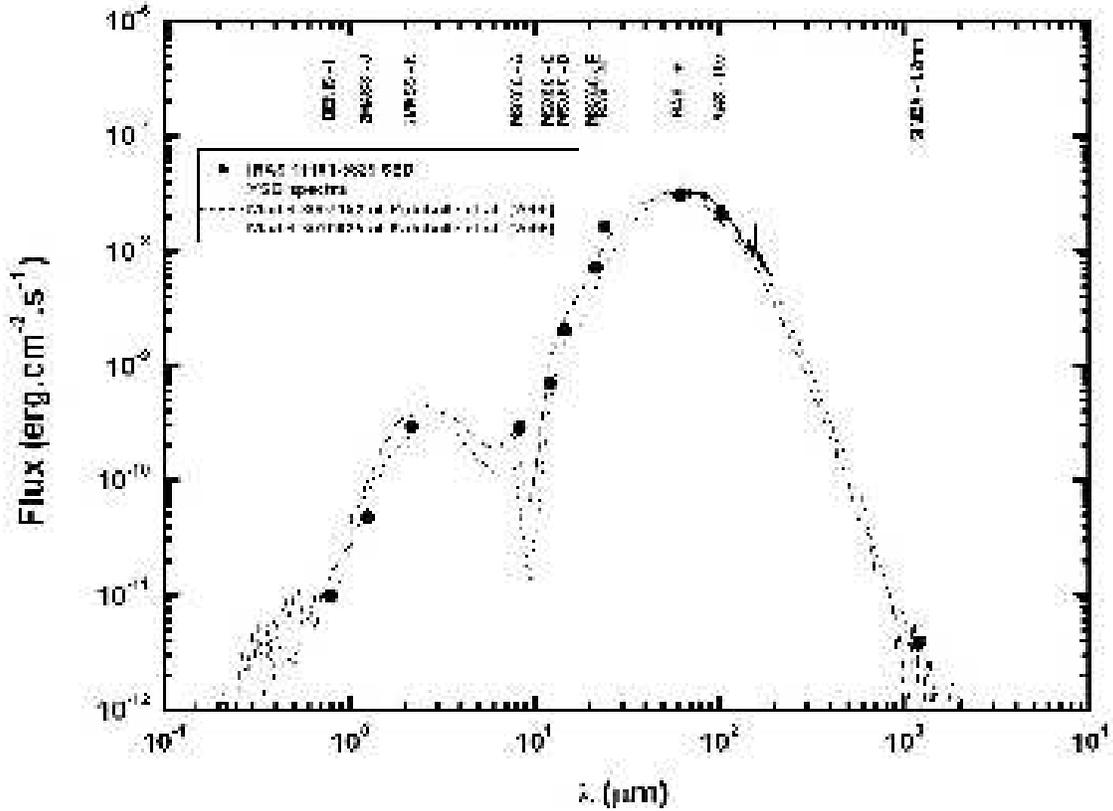}
\caption{Spectral energy distribution of IRAS 11101-5829. 
The observational data are from
the following resources. DENIS: \citet{denis}. 2MASS: \citet{cut03}.
MSX: \citet{ega03}. IRAS: \citet{bec88}.
 SIMBA: \citet{hil05}.   ISO spectrum
from the ISO data archive (http://www.iso.vilspa.esa.es/ida/index.html).
The dotted and traced lines represent two models from \citet{rob06} (see
text for details).}
\label{fig_sed}
\end{figure}

CVR would like to thank J. W. Vilas-Boas for fruitful discussions.
We acknowledge the use of: the USNOFS Image and Catalogue Archive
operated by the United States Naval Observatory, Flagstaff Station
(http://www.nofs.navy.mil/data/fchpix/); the SIMBAD database, operated at
CDS, Strasbourg, France;  the VizieR catalogue access tool, CDS, Strasbourg,
France; the NASA's Astrophysics Data System Service;
and NASA's {\it SkyView} facility (http://skyview.gsfc.nasa.gov) located 
at NASA Goddard Space Flight Center. 
Use of the images in Figures \ref{fig_small}, \ref{fig_largep}, and \ref{fig_large_scale}
is courtesy
of the UK Schmidt Telescope (copyright in which is owned by the Particle 
Physics and Astronomy Research Council of the UK and the Anglo-Australian 
Telescope Board) and the Digitized Sky Survey
created by the Space Telescope Science Institute, operated by AURA, Inc, for
NASA, and is reproduced here with permission from the Royal Observatory
Edinburgh. This work was partially supported by Fapesp (CVR: Proc. 2001/12589-1).



{\it Facilities:} \facility{LNA:BC0.6m ()}




\clearpage

\clearpage

\clearpage

\clearpage

\clearpage

\clearpage

\clearpage

\clearpage

\clearpage

\clearpage

\clearpage

\begin{table}
\begin{center}
\caption{Parameters of the Gaussian fits to polarimetric data in the line-of-sight
to \pairhh
\label{tab-gauss}}
\begin{tabular}{ccccccc}
\tableline\tableline
Suggested origin & Mean & Error & Standard  & Error & Data \\
& & &  Deviation \\
 & (\degr) & (\degr) & (\degr) & (\degr)\\
\tableline
\pairhh region - with foreground & 54.9 & 1.4 & 14.2 & 1.0 & this work \\
\pairhh region - foreground subtracted & 41.9 & 1.2 & 13.8 & 1.0 & this work \\
foreground ISM - 10\arcmin $\times$ 10\arcmin\ field & 100.5 & 0.4 & 10.2 & 0.3 & this work\\
foreground ISM - 5\degr $\times$ 5\degr\ field & 107.7 & 2.3 & 24.8 & 2.9 & \citet{hei00} \\
\tableline
\end{tabular}
\end{center}
\end{table}

\clearpage

\begin{table}
\begin{center}
\caption{Parameters of \citet{rob06} models that best fit the \iras SED
\label{tab-sed}}
\begin{tabular}{ccccccc}
\tableline\tableline
Parameter Description & Value  \\
\tableline
Model \# & 3007152 & 3020025\\
\tableline
Stellar Mass (\mmsol) 	& 12.87 & 13.01\\
Stellar Radius (R$_\odot$) 	& 4.47 & 4.49 \\
Stellar Temperature (K) 	& 29,390 & 29,545\\
Envelope Accretion Rate (\mmsol/yr) 	& $1.04\times10^{-3}$ & $6.71\times10^{-4}$ \\
Envelope Outer Radius (AU) 	& $1.00\times10^{5}$ & $1.00\times10^{5}$\\
Cavity Angle (degrees) 	& 31.8 & 29.4 \\
Disk Mass (\mmsol) 	& $3.64\times10^{-2}$ & $1.69\times10^{-2}$ \\
Disk Outer Radius (AU) 	& 86.6 & 159\\
Disk Inner Radius (Rsub) 	& 1.00 & 1.00 \\
Disk Inner Radius (AU) 	& 9.51 & 9.53 \\
Scaleheight factor 	& 0.837& 0.996 \\
Disk Flaring Power 	& 1.073 & 1.104 \\
Cavity density (cgs) 	& $1.25\times10^{-20}$ & $1.12\times10^{-20}$\\
Ambient density (cgs) 	& $6.09\times10^{-21}$ & $6.54\times10^{-21}$\\
Disk accretion rate (\mmsol/yr) 	& $5.72\times10^{-6}$ & $7.04\times10^{-7}$ \\
Total A(V) along line-of-sight 	& $4.90\times10^{3}$ & $0.99\times10^{3}$ \\
Total Luminosity (\llsol) 	& $1.38\times10^{4}$ & $1.38\times10^{4}$ \\
Disk Scaleheight at 100AU 	& 6.84 & 8.77 \\
\tableline
\end{tabular}
\end{center}
\end{table}


\begin{thebibliography}{}

\bibitem[Bastien et al.(1988)]{bas88}Bastien, P., Drissen, L., M\'enard, F., Moffat,
A.~F.~J., Robert, C., \& St-Louis, N.\ 1988, AJ, 95, 900

\bibitem[Beichman et al.(1988)]{bec88} Beichman, C.~A., 
Neugebauer, G., Habing, H.~J., Clegg, P.~E., \& Chester, T.~J.\ 1988, 
Infrared astronomical satellite (IRAS) catalogs and atlases.~Volume 1: 
Explanatory supplement, 1,  

\bibitem[Bernasconi \& Maeder(1996)]{ber96} Bernasconi, 
P.~A., \& Maeder, A.\ 1996, \aap, 307, 829 

\bibitem[Brand et al.(1986)]{bra86}Brand, J., Blitz, L., \& Wouterloot,
J.~G.~A.\ 1986,
A\&AS, 65, 537

\bibitem[Braz et al.(1989)]{bra89}Braz, M.~A., Gregorio-Hetem, J.~C., Scalise,
E., Jr.,
Monteiro Do Vale, \& J.~L., Gaylard, M.\ 1989, A\&AS, 77, 465

\bibitem[Bronfman et al.(1996)]{bro96}Bronfman, L., Nyman, \& L.-A., May, J.\
1996, A\&AS, 115, 81

\bibitem[Brooks et al.(2003)]{bro03}{Brooks, K.~J., Garay, G., Mardones, D., \&
Bronfman, L.\ 2003, \apjl, 594, L131}

\bibitem[Chandrasekhar \& Fermi(1953)]{cha53}Chandrasekhar, S. \& Fermi, E.\
1953,
ApJ, 118, 113

\bibitem[Chrysostomou et al.(1994)]{chr94} Chrysostomou, A., 
Hough, J.~H., Burton, M.~G., \& Tamura, M.\ 1994, \mnras, 268, 325 

\bibitem[Chrysostomou et al.(2006)]{chr06}Chrysostomou, A., Lucas, P.~W.,
Hough, J.H. \& Tamura, M.\ 2006, in Protostars and Planets 2005, Eds. B.
Reipurth, D. Jewitt, \&
K. Keil (Tucson: University of Arizona Press), in press

\bibitem[Cutri et al.(2003)]{cut03} Cutri, R.~M., et al.\ 
2003, The IRSA 2MASS All-Sky Point Source Catalog, NASA/IPAC Infrared 
Science Archive.~http://irsa.ipac.caltech.edu/applications/Gator/,  

\bibitem[Davis et al.(2004)]{dav04}{Davis, C.~J., Varricatt, W.~P., Todd, S.~P.
\& Ramsay Howat, S.~K.\ 2004, \aap, 425, 981}

\bibitem[Davis \& Greenstein(1951)]{dav51} Davis, L. \& Greenstein, J.~L.\ 1951,
ApJ, 114, 206

\bibitem[De Colle \& Raga(2005)]{dec05}{De Colle, F. \& Raga, A.~C.\ 2005,
\mnras, 359, 164}

\bibitem[Dopita(1978)]{dop78} Dopita, A.\ 1978, \aap, 63, 237

\bibitem[Dutra et al.(2003)]{dut03}Dutra, C. M., Bica, E., Soares, J., \&
Barbuy, B.
2003, A\&A, 400, 533

\bibitem[Egan et al.(2003)]{ega03} Egan, M.~P., et al.\ 2003, 
VizieR Online Data Catalog, 5114, 0 

\bibitem[Ferreira et al.(2006)]{fer06}Ferreira, J., Dougados, C., \& Cabrit, S.\
2006, A\&A, 453, 785

\bibitem[Gonatas et al.(1990)]{gon90}Gonatas, D.~P., Engargiola, G.~A.,
Hildebrand, R.~H., Platt, S.~R., Wu, X.~D., Davidson, J.~A., Novak, G.,
Aitken, D.~K. \& Smith, C.\ 1990, ApJ, 357, 132

\bibitem[Gredel(2006)]{gre06}Gredel, R.\ 2006, A\&A, 457,157

\bibitem[Hartley et al.(1986)]{har86}Hartley, M., Manchester, R.~N., Smith,
R.~M., Tritton, S.~B., \& Goss, W.~M.\ 1986, A\&AS, 63, 27

\bibitem[Heiles(2000)]{hei00}Heiles, C.\ 2000, \aj, 119, 923

\bibitem[Heitsch(2005)]{hei05}Heitsch, F.\ 2005, ASP 
Conf.~Ser.~343: Astronomical Polarimetry: Current Status and Future 
Directions, 343, 166 

\bibitem[Heitsch et al.(2001)]{hei01}Heitsch, F., Zweibel, E.~G., Mac Low,
M.-M., Li, P., \& Norman, M.~L.\ 2001, ApJ, 561, 800

\bibitem[Hill et al.(2005)]{hil05}Hill, T., Burton, M.~G., 
Minier, V., Thompson, M.~A., Walsh, A.~J., Hunt-Cunningham, M., \& Garay, 
G.\ 2005, \mnras, 363, 405 

\bibitem[Lazarian(2003)]{laz03} Lazarian, A.\ 2003, Journal 
of Quantitative Spectroscopy and Radiative Transfer, 79, 881 

%
\bibitem[Magalh\~aes et al.(1984)]{mag84}Magalh\~aes, A.~M., Benedetti, E., \&
Roland, E., 1984, PASP, 96, 383

\bibitem[Magalh\~aes et al.(1996)]{mag96}Magalh\~aes, A.~M., Rodrigues, C.~V.,
Margoniner, V.~E., Pereyra, A.,  \& Heathcote, S., 1996, in ASP Conf. Ser. 97,
Polarimetry of the Interstellar Medium, Eds. W.~G. Roberge \& D.~C.~B. Whittet 
(San Francisco:ASP), 118

\bibitem[Mart{\'{\i}} et al.(1993)]{mar93}{Mart{\'{\i}}, J., 
Rodr{\'{\i}}guez, L.~F., \& Reipurth, B.\ 1993, \apj, 416, 208}

\bibitem[Matsumoto et al.(2006)]{mat06} Matsumoto, T., 
Nakazato, T., \& Tomisaka, K.\ 2006, \apjl, 637, L105 

\bibitem[M\'enard \& Duch\^ene(2004)]{men04}M\'enard, F. \& Duch\^ene, G.\ 2004,
A\&A, 425, 973

\bibitem[Ogura et al.(1998)]{ogu98}Ogura, K., Nakano, M., Sugitani, K., \&
Liljestr\"om, T. 
1998, A\&A, 338, 576

\bibitem[Ogura \& Walsh(1992)]{ogu92}Ogura, K. \& Walsh, J.~R.\ 1992, ApJ, 400,
248

\bibitem[Ostriker et al.(2001)]{ost01}Ostriker, E.~C., Stone, J.~M., \& Gamie,
C.~F.\ 2001, ApJ, 546, 980

\bibitem[Otrupcek et al.(2000)]{otr00}Otrupcek, R.~E., Hartley, M., \& Wang,
J.-S.
2000, PASA, 17, 92

\bibitem[Padoan et al.(2001)]{pad01}Padoan, P., Goodman, A., Draine, B.~T.,
Juvela, M., 
Nordlund, \AA, \& R\"ognvaldsson, \"O.~E.\ 2001, ApJ, 559, 1005

\bibitem[Pereyra(2000)]{per00}Pereyra, A., 2000. Dust and Magnetic Fields in
Dense Regions of the
Interstellar Medium, PhD Thesis, Univ. S\~ao Paulo

\bibitem[Pereyra \& Magalh\~aes(2005)]{per05}Pereyra, A., \& Magalh\~aes, A.~M.\
2005,
in AIP Conf. Proc. 784, Magnetic Fields in the Universe, ed. E.~M. de Gouveia
dal Pino, G. Lugones,
\& A. Lazarian (Melville: American Institute of Physics), 743

\bibitem[Piirola(1973)]{pii73} Piirola, V., 1973, A\&A, 27, 383

\bibitem[Purcell \& Spitzer(1971)]{pur71}Purcell, E.~M., \& 
Spitzer, L.~J.\ 1971, \apj, 167, 31 

\bibitem[Robitaille et al.(2006)]{rob06} Robitaille, T.~P., 
Whitney, B.~A., Indebetouw, R., Wood, K., \& Denzmore, P.\ 2006, \apjs, 
167, 256 

\bibitem[Serkowski et al.(1975)]{ser75}Serkowski, K., Mathewson, D.~L., \&
Ford, V.~L.\ 1975, ApJ, 196, 261

\bibitem[Shang et al.(2006)]{sha06}Shang, H., Li, Z.-Y., \& Hirano, N.\ 2006, 
in Protostars and Planets 2005, Eds. B. Reipurth, D. Jewitt, \&
K. Keil (Tucson: University of Arizona Press), in press

\bibitem[Tamura et al.(1997)]{tam97}Tamura, M., Hough, J.H., Chrysostomou, A.,
Itoh, Y., Murakawa, K., \& Bailey, J.~A.\ 1997, MNRAS, 287, 894

\bibitem[Te Lintel Hekkert \& Chapman(1996)]{tel96}Te Lintel Hekkert, P., \& 
Chapman, J.~M.\ 1996, A\&AS, 119, 459

\bibitem[The Denis Consortium(2005)]{denis}The Denis 
Consortium 2005, VizieR Online Data Catalog, 1, 2002 

\bibitem[Turnshek et al.(1990)]{tur90}Turnshek, D.~A., Bohlin, R.~C., Williamson,
R.~L. II, Lupie, O.~L.,  Koornneef, J., \& Morgan, D. H.\ 1990, AJ, 99, 1243

\bibitem[Vrba et al.(1988)]{vrb88} Vrba, F.~J., Strom, S.~E., 
\& Strom, K.~M.\ 1988, \aj, 96, 680 

\bibitem[Walsh et al.(1997)]{wal97}Walsh, A.~J., Hyland, A.~R., Robinson, G., 
\& Burton, M.~G.\ 1997, MNRAS, 291, 261

\bibitem[Zinchenko et al.(1995)]{zin95}Zinchenko, I., Mattila, K., \& Toriseva,
M.\ 1995, 
A\&AS, 111, 95

%
\bibitem[Zweibel(1996)]{zwe96}Zweibel, E.~G.\ 1996, in ASP Conf. Ser. 97,
Polarimetry of the Interstellar Medium, Eds. W.~G. Roberge \& D.~C.~B. Whittet 
(San Francisco:ASP), 486


\end{thebibliography}
\end{document}